\documentclass[conference]{IEEEtran}
\usepackage{alltt}
\usepackage{amssymb}
\usepackage{graphicx}
\usepackage{url}
\usepackage{version}

\includeversion{LONG}
\excludeversion{SHORT}

\ifCLASSINFOpdf
\else
\fi
\newtheorem{fact}{Fact}
\newtheorem{lemma}{Lemma}
\newtheorem{property}{Property}
\newtheorem{theorem}{Theorem}

\newcommand{\secref}[1]{\S\,\ref{#1}}

\begin{document}
%
\begin{LONG}
\title{Automated Validation of Security-sensitive Web Services
  specified in BPEL and RBAC \\ {\Large (Extended Version)}}
\end{LONG}

\begin{SHORT}
\title{Automated Validation of Security-sensitive Web Services
  specified in BPEL and RBAC}
\end{SHORT}

\author{\IEEEauthorblockN{Alberto Calvi}
\IEEEauthorblockA{Dipartimento di Informatica\\Universit\`a di Verona, Italy\\
Email: alberto.calvi@univr.it}
\and
\IEEEauthorblockN{Silvio Ranise}
\IEEEauthorblockA{FBK-Irst (Trento, Italy)\\
Email: ranise@fbk.eu}
\and
\IEEEauthorblockN{Luca Vigan\`o}
\IEEEauthorblockA{Dipartimento di Informatica\\Universit\`a di Verona, Italy\\
Email: luca.vigano@univr.it}}


%


\maketitle

\begin{abstract}
  We formalize automated analysis techniques for the validation of web
  services specified in BPEL and a RBAC variant tailored to BPEL.
  The idea is to use decidable fragments of first-order logic to
  describe the state space of a certain class of web services and then
  use state-of-the-art SMT solvers to handle their reachability
  problems.
To assess the practical viability of our approach, we have developed a prototype tool implementing our techniques and applied it to a digital contract signing service inspired by an industrial case study.
\end{abstract}


%
\IEEEpeerreviewmaketitle

\section{Introduction}

\paragraph*{Context and motivation} 

The design of security-sensitive web services is an error-prone and
time-consuming task. The reasons of these difficulties are manyfold. A
web service is (often) obtained as a composition of several simpler
services executed in a distributed environment. So, because of the huge
number of possible interleavings and the subtle interplay between the
data and the control part of the processes, it is very difficult---if
not impossible---for a human to foresee all the possible behaviors.
Furthermore, the workflow of the application is usually constrained by
the enforcement of access-control policies that forbid the execution of
certain operations or the access to shared resources by certain users 
and can easily over-constrain or
under-constrain the possible behaviors. As a consequence, in the first
case, correct behaviors are prevented, thereby decreasing the overall
dependability of the service, while in the second case, incorrect
behaviors are possible that may open security breaches, thus destroying
the dependability of the service.

Hence, automated techniques for the validation of security-sensitive web
services at design time are needed to assist the designers and avoid
expensive actions for the correction of errors after deployment. While
this is only a preliminary step in the direction of building highly
dependable web services, it constitutes a necessary stepping stone for
the application of other techniques at run-time for orchestration and
coordination of services and for enforcing access policies.

\paragraph*{Contributions} 

In this paper, we formalize automated analysis techniques for the
validation of web services specified in BPEL and a variant of RBAC
tailored to BPEL as proposed in~\cite{bertino-etal}. RBAC (see,
e.g.,~\cite{rbac}) is one of the most successful models for access
control in large and complex applications. Our idea is to translate a
BPEL process to a particular class of transition systems, described by
arithmetic constraints and called Vector Addition Systems (VASs), and to
encode the RBAC specification in a decidable class of first-order
formulae, called Bernays-Sch\"onfinkel-Ramsey (BSR). We study the goal
reachability problem (to which several analysis problems can be reduced)
of the resulting class of transition systems. Theoretically, we prove
the decidability of the reachability problem for a particular class of
transition systems modeling BPEL processes where no loops occur.
Pragmatically, to assess the viability of our approach, we have
developed a prototype tool called WSSMT, which implements our techniques
and uses state-of-the-art theorem-proving techniques recently developed
in the area of Satisfiability Modulo Theories (SMT) and featuring a good
trade-off between scalability and expressiveness. We report on the
application of WSSMT on a digital contract signing service inspired by
an industrial case study.

\paragraph*{Related work}  
While BPEL semantics is given in natural language in~\cite{bpel}, there
have been many attempts to give a formal semantics of the language in
terms of Petri nets, e.g.,~\cite{stahl}. The formalization is useful in
two respects: it eliminates possible ambiguities in the natural language
semantics and 
it permits the formal
analysis of BPEL processes at the design time.
Although there are tools
(e.g.,~\cite{bpel2owfn}) that provide automated support for the
translation from BPEL to Petri nets and the subsequent analysis, they
(to the best of our knowledge) only model the control flow and abstract
away from the data manipulation. Recently, there have been attempts at
extending Petri nets with some data modeling and reasoning capabilities
by using fragments of first-order logic (FOL) for which efficient SMT solvers
exist~\cite{monakova}. Instead of a hybrid representation, we chose to
develop our techniques in a first-order framework by exploiting the
well-known connection between Petri nets and VASs (see,
e.g.,~\cite{reachability-timed-petri-net,manna-petri-net-invariant-generation}) and
then to extend it along the lines suggested in~\cite{passat09} to
incorporate the access-control layer in a uniform way.
The work in~\cite{passat09} studies the decidability of symbolic
executions with bounded length for more general classes of services
while here we focus on a particular class of applications whose sets
of reachable states can be finitely described by suitable fragments of
FOL.
The work in~\cite{fast09-armando} is closely related to ours with
respect to the structure of the specification divided in two layers,
one for the workflow and one for access
control. However,~\cite{fast09-armando} does not provide a
decidability result for the reachability problem as we do in this
paper.

\paragraph*{Organization} \secref{sec:motivating} briefly introduces the
languages used to specify the class of web services we consider, i.e.\
BPEL and RBAC4BPEL, together with a concrete example that we use to
illustrate the key features of the formal framework.
\secref{sec:formalization} recalls the definition of two-level
transition system introduced in~\cite{passat09}, and its related
reachability problem, explains how BPEL and RBAC4BPEL specifications can
be translated to this class of transition systems, and proves the
decidability of the reachability problem for two-level transition
systems obtained by translating a class of acyclic Petri nets (called
workflow nets). \secref{sec:industrial} discusses how our techniques
have been implemented and applied to a digital contract signing service,
inspired by an industrial application. 
In \secref{sec:conclusion}, we draw conclusions and discuss future work. 
\begin{LONG}
Proofs of the formal results are given in an appendix.
\end{LONG}
\begin{SHORT}
Proofs of the formal results are given in~\cite{CRV10-long}.
\end{SHORT}

\section{BPEL, RBAC4BPEL, and a motivating example}
\label{sec:motivating}
We characterize the class of applications we are interested in by using
the Purchase Ordering (PO) process introduced in~\cite{bertino-etal}. To
make the paper self contained, in this section, we briefly illustrate
the example and give a high-level description of the languages used to
specify it.
%
The PO process is composed of six activities: the
creation of a purchase order for a certain good (crtPO), the approval
of the order before dispatching the good to the supplier (apprPO), the
acknowledgement of the delivery by signing (signGRN) and then
countersigning (ctrsignGRN) the goods-received note, the creation of a
payment file on receipt of the supplier's invoice for the good
(crtPay), and the approval of the payment to the supplier (apprPay).  
For the PO process to complete successfully, the order of execution of
the various activities should satisfy the following constraints: crtPO
must be executed before apprPO which, in turn, must be executed before
the remaining four activities; crtPay can be done in parallel with both
signGRN and ctrsignGRN but before apprPay; and signGRN, ctrsignGRN, and
apprPay must be executed in this order. 
The \emph{workflow (WF) level} of the
application should enforce these dependencies that are induced by the
application logic of the PO process.  


\subsection{The WF level and BPEL}

\begin{figure}
	\begin{minipage}{.4\textwidth}\footnotesize
      \begin{alltt}
<process name="PO"/>
  <sequence>
    <receive operation="crtPO" ... > </receive>
    <invoke  operation="apprPO" ... > </invoke>    
    <flow>
      <sequence>
        <invoke operation="signGRN" ... > </invoke>
        <invoke operation="ctrsignGRN" ... > </invoke>
      </sequence>
      <invoke operation="ctrPay" ... > </invoke>
    </flow>
    <invoke operation="apprPay" ... > </invoke>
  </sequence>
</process>
      \end{alltt}
    \end{minipage}
	\caption{\label{fig:ex-forms-bpel}The WF level of the PO process: BPEL.}
\end{figure}

\begin{figure}[ht]
	\begin{minipage}{.4\textwidth}
		\includegraphics[height=.5\textheight]{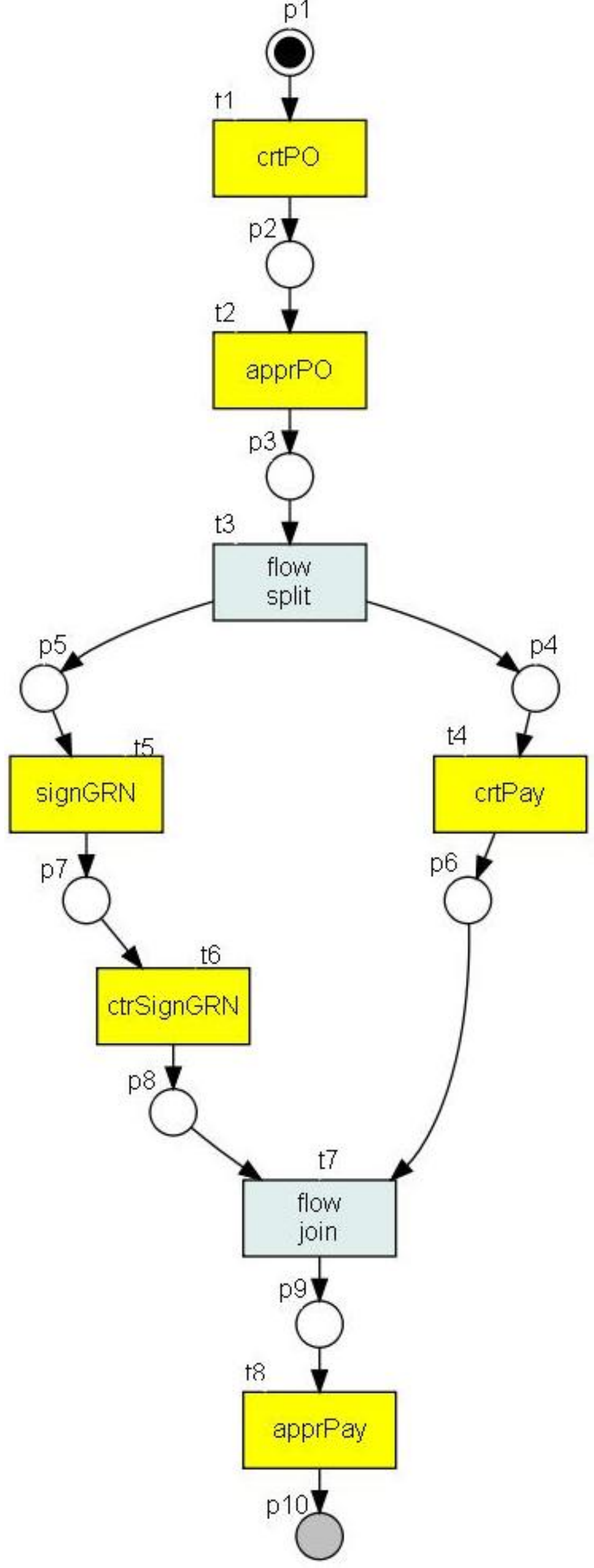} 
	\end{minipage} 
	\caption{\label{fig:ex-forms-pn}The WF level of the PO process: Petri Net corresponding
	to BPEL in Fig.~\ref{fig:ex-forms-bpel}.}
\end{figure}

\begin{figure*}[ht]
	\begin{eqnarray*}
		\begin{array}{l}
			U ~ := ~ \{ u_1, u_2, u_3, u_4, u_5 \} \qquad
			R ~ := ~ \{ Manager, FinAdmin, FinClerk, POAdmin, POClerk\} \\
			P ~ := ~ \{ p_1, ..., p_5\} \qquad
			ua ~ := ~ \{ (u_1,Manager), (u_2,FinAdmin), (u_3,FinClerk), 
									(u_4,POAdmin), (u_5,POClerk) \} \\
			pa ~ := ~ \{ (FinClerk,p_4), (FinAdmin,p_5), 
									(POClerk,p_3), (POAdmin,p_1)   \} \\
			\succeq \mbox{ least partial order s.t. } 
			Manager \succeq FinAdmin, ~Manager \succeq POAdmin, \\
			~~~~~~~~~~~~~~~~~~~~~~~~~~~~~~ 
			FinAdmin \succeq FinClerk, \mbox{ and } POAdmin \succeq POClerk .
		\end{array}
  \end{eqnarray*}
	
	\caption{\label{fig:ex-forms-rbac}The PM level of the PO process: RBAC}
\end{figure*}

In~\cite{bertino-etal}, the Business Process Execution Language
(BPEL~\cite{bpel}) is used to describe a (executable) specification
of the WF level of an application.  
In Fig.~\ref{fig:ex-forms-bpel}, we show a high-level BPEL
specification of the WF level of the PO process. The \texttt{<process>}
element wraps around the entire description of the PO process. The
\texttt{<sequence>} element states that the activities contained in its
scope 
must be executed sequentially. The \texttt{<flow>} element specifies
concurrent threads of activities. The \texttt{<invoke>} element
represents the invocation of an activity that is provided by an
available web service. Finally, the \texttt{<receive>} element
represents the invocation of an activity that is provided by the BPEL
process being described. Indeed, BPEL provides a variety of constructs
(e.g., to represent variables) that are ignored here for simplicity;
the interested reader is pointed to~\cite{bpel}. In the case of the PO
process, it is easy to see that the constraints on the execution
described above are all satisfied by the nesting of control elements
in Fig.~\ref{fig:ex-forms-bpel}. For example, because of the
semantics of \texttt{<sequence>}, crtPO will be executed first while
apprPay will be the activity finishing the PO process.

Fig.~\ref{fig:ex-forms-pn} shows a
Petri net that can be seen as the (formal) semantics of the BPEL
process in Fig.~\ref{fig:ex-forms-bpel}.  
Before being able to sketch the mapping from BPEL processes to Petri
nets, we recall the basic notions concerning the latter.

A \emph{Petri net} is a triple $\langle P, T, F \rangle$, where $P$ is
a finite set of \emph{places}, $T$ is a finite set of \emph{transitions}, and
$F$ (\emph{flow relation}) is a set of arcs such that $P\cap
T=\emptyset$ and $F\subseteq (P\times T)\cup (T\times
P)$. Graphically, the Petri net $\langle P, T, F \rangle$ can be
depicted as a directed bipartite graph with two types of nodes,
places and transitions, represented by circles and
rectangles, respectively; the nodes are connected via directed arcs
according to $F$ (where arcs between two nodes of the same type are
not allowed). A place $p$ is called an \emph{input} (resp.,
\emph{output}) place of a transition $t$ iff there exists a directed
arc from $p$ to $t$ (resp., from $t$ to $p$). The set of input (resp.,
output) places of a transition $t$ is denoted by $\bullet t$ (resp.,
$t\bullet$);
$\bullet p$ and $p\bullet$ are defined similarly.  
A \emph{path} in a Petri net $\langle P, T, F \rangle$ is a finite
sequence $e_0, ..., e_n$ of elements from $P\cup T$ such that
$e_{i+1}\in e_i\bullet$ for each $i=0, ..., n-1$; a path $e_0,...,
e_n$ in the net is a \emph{cycle} if no element occurs more than once
in it and $e_0\in e_n\bullet$ for some $n\geq 1$. A Petri net is
\emph{acyclic} if none of its paths is a cycle. A \emph{marking} of a
Petri net $\langle P, T, F \rangle$ is a mapping from the set $P$ of
places to the set of non-negative integers; graphically, it is
depicted as a distribution of black dots in the circles of the graph
representing the net. A transition $t$ is \emph{enabled} in a marking
$m$ iff each of its input places $p$ is such that $m(p)\geq 1$, i.e.,
$p$ contains at least one token. An enabled transition $t$ in a
marking $m$ may \emph{fire} by generating a new marking $m'$, in
symbols $m\stackrel{t}{\rightarrow} m'$, where $m'(p)=m(p)$ if
$p\not\in (\bullet t\cup t\bullet)$, $m'(p)=m(p)-1$ if $p\in \bullet
t$, and $m'(p)=m(p)+1$ if $p\in t \bullet$, i.e.\ $t$ consumes one
token from each input place of $t$ and produces one token in each of
its output places. A marking $m$ is \emph{reachable} from $m_0$, in
symbols $m_0\rightarrow^* m$, iff there exists a sequence $m_1, ...,
m_n$ of markings such that $m_i\stackrel{t}{\rightarrow} m_{i+1}$ for
$i=0, ..., n-1$ and $m_n=m$, for some $n\geq 0$. (In case $n=0$, we
have that $m_0=m$.) Given a Petri net $\langle P, T, F \rangle$ and a
marking $m$, an instance of the \emph{reachable problem} for Petri
nets consists of checking whether $m_0\rightarrow^* m$ or not. A
\emph{workflow (WF) net}~\cite{wn-vanderAlst} is a Petri net $\langle P,
T, F \rangle$ such that (a) there exist two special places $i,o\in P$
with $\bullet i=\emptyset$ and $o\bullet=\emptyset$; and (b) for each
transition $t\in T$, there exists a path $\pi$ in the net beginning with
$i$ and ending with $o$ in which $t$ occurs.

The idea underlying the Petri net semantics of BPEL is simple.
Activities are mapped to transitions (the rectangles in
Fig.~\ref{fig:ex-forms-pn}) and their execution is modeled by the flow of tokens from input
places to output places. When two BPEL operations are enclosed in a
\texttt{<sequence>} element (e.g., crtPO and apprPO), two transitions
are created (as in Fig.~\ref{fig:ex-forms-pn}) with one input 
place (resp., $p_1$ and $p_2$) and one output place each (resp., $p_2$ 
and $p_3$), and the input place of the second is identified 
with the output place of the first one ($p_2$).
When two BPEL operations are in a \texttt{<flow>} element (e.g.,
ctrPay and the sequence of operations signGRN and ctrsignGRN), four
transitions are created: one to represent the split of the flow, one
to represent its synchronization (join), and one
for each activity that can be executed concurrently with the
appropriate places to connect them
(in Fig.~\ref{fig:ex-forms-pn}, when a token is in place $p_3$, the
`flow split' transition is enabled and its execution yields one token
in place $p_4$ and one in $p_5$, which enables both transitions signGRN
and ctrPay that can be executed concurrently; the two independent
threads of activities get synchronized again by the execution of `flow
join', which is enabled when a token is in $p_6$ and a token is in
$p_8$). 
It is easy to see that the Petri net of Fig.~\ref{fig:ex-forms-pn} is an
acyclic WF net where $p_1$ is the special input place $i$, 
$p_{10}$ is the special output place $o$, and each transition occurs in
a path from $p_1$ to $p_{10}$.

\subsection{The policy management (PM) level and RBAC4BPEL}
\label{subsec:rbac4bpel}

Besides the dependencies imposed by the WF level, constraints on the
execution of the activities derived from security requirements are of
crucial importance to ensure the dependability of the application. In
this paper, we focus on a particular class of security requirements that
pertain to the access-control policy. The \emph{policy management (PM)
level} of the application is charged to enforce these constraints.

In~\cite{bertino-etal}, an extension of the Role Based Access Control
(RBAC) framework---adapted to work smoothly with BPEL, denoted with
RBAC4BPEL---is used to specify the PM level of applications. The
components of RBAC4BPEL are: (i) a set $U$ of users, (ii) a set $R$ of
roles, (iii) a set $P$ of permissions, (iv) a role hierarchy $\succeq$
(i.e.\ a partial-order relation on $R$), (v) a user-role assignment
relation $ua$, (vi) a role-permission assignment relation $pa$, and
(vii) a set $A$ of activities and a class of authorization constraints
(such as separation-of-duty) to prevent some user to acquire
permissions in certain executions of the application (see below for
details). Note that components (i)--(vi) are standard in RBAC while
(vii) has been added to obtain a better integration between the PM and
the WF levels.

First, we describe components (i)--(vi) and some related notions. A user
$u\in U$ is assigned a role $r\in R$ if $(u,r)\in ua$ and permissions
are associated with roles when $(p,r)\in pa$. In RBAC4BPEL, a user $u\in
U$ has a permission $p$ if there exists a role $r\in R$ such that
$(u,r)\in ua$ and $(p,r)\in pa$. (We will see that each permission is
associated to a right on a certain activity in $A$---e.g., its
execution---of a BPEL process.) The role hierarchy $\succeq\subseteq
R\times R$ is assumed to be a partial order (i.e., a reflexive,
antisymmetric, and transitive relation) reflecting the rights associated
to roles. More precisely, a user $u$ is an \emph{explicit} member of
role $r\in R$ if $(u,r)\in ua$ and it is an \emph{implicit} member of
role $r\in R$ if there exists a role $r'\in R$ such that
$(r',r) \in \succeq$ 
(abbreviated as $r'\succeq r$), $r'\neq r$, and $(u,r')\in ua$.  Thus,
$\succeq$ induces a permission inheritance relation
as follows: a user $u\in U$ \emph{can get} permission $p$ if there
exists a role $r\in R$ such that $u$ is a member (either implicit or
explicit) of $r$ and $(p,r)\in pa$. For simplicity, we abstract away
the definition of a role in terms of a set of attributes as done
in~\cite{bertino-etal}.

Fig.~\ref{fig:ex-forms-rbac} shows the sets $U, R, P$ and the
relations $\succeq, ua, pa$ for the PM level of the PO process. Although
$(Manager,p_i)\not \in pa$ for any $i=1,...,5$, we have that user $u_1$,
which is explicitly assigned to role $Manager$ in $ua$, can get 
%
%
%
%
any permission $p_i$ for $i=2, ...,5$ as $Manager\succeq r$ 
for any role $r\in R\setminus\{Manager\}$, hence $u_1$ can be
implicitly assigned to each role and then get the permission $p_i$.

In RBAC4BPEL, each permission in $P$ is associated with the right to
handle a certain transition of $T$, uniquely identified by a label in
$A$, for a Petri net $\langle P, T, F \rangle$. For the PO process,
this is particularly simple since only the right to execute a
transition is considered. So, $p_1$ is the permission
for executing apprPO, $p_2$ for signGNR, $p_3$ for ctrSignGNR, $p_4$
for crtPay, and $p_5$ for apprPay. We are now in the position to
describe component (vii) of RBAC4BPEL. Note that there are no
permissions associated to `flow split' and `flow join' as these are
performed by the BPEL engine and thus no particular authorization
restriction must be enforced.

A \emph{role} (resp., \emph{user}) \emph{authorization constraint} is
a tuple $\langle D, (t_1,t_2), \rho \rangle$ if $D\subseteq R$ (resp.,
$D\subseteq U$) is the domain of the constraint, $\rho\subseteq
R\times R$ (resp., $\rho\subseteq U\times U$), and $t_1,t_2$ are in
$A$.  An authorization constraint $\langle D, (t_1,t_2), \rho \rangle$
is \emph{satisfied} if $(x,y)\in \rho$ when $x,y\in D$, $x$ performs
$t_1$, and $y$ performs $t_2$. In other words, authorization
constraints place further restrictions (besides those of the standard
RBAC components) on the roles or users who can perform certain actions
once others have been already executed by users belonging to certain
roles. Constraints of this kind allow one to specify
separation-of-duty (SoD) by $\langle D, (t_1,t_2), \neq \rangle$,
binding-of-duty (BoD) by $\langle D, (t_1,t_2), = \rangle$, or any
other restrictions that can be specified by a binary relation over
roles or users.

For the PO process, (vii) of RBAC4BPEL is instantiated as:

\vspace*{-0.4cm}
{\footnotesize
\begin{eqnarray*}
  \langle U, (apprPO, signGNR), \neq \rangle, \ 
  \langle U, (apprPO, crtSignGNR), \neq \rangle, \\
  \langle U, (signGNR, crtSignGNR), \neq \rangle, \
  \langle R, (crtPay, apprPay),   \prec \rangle, 
\end{eqnarray*}
}%
where $\prec := \{ (r_1, r_2) ~|~ r_1, r_2\in R,~ r_2\succeq r_1,~
r_1\neq r_2\}$ (recall that the sets $U$ and $R$ are defined in
Fig.~\ref{fig:ex-forms-rbac}).

This concludes the description of the class of applications that we
consider.  We now proceed to introduce our techniques to analyze such
applications.

\section{Formalization and Automated Analysis}
\label{sec:formalization}

From now on, we assume that the WF level of an application is
specified by a Petri net and the PM level by an instance of the
RBAC4BPEL framework. 
We use 
two-level transition systems~\cite{passat09} to represent the WF level
and the PM level of a web-service and we study the reachability
problem for a sub-class.

\subsection{Two-level transition systems and goal reachability}

We assume the basic notions of FOL (see,
e.g.,~\cite{enderton}). A \emph{two-level transition system} $Tr$ is a
tuple

\vspace*{-0.4cm}
{\footnotesize
\begin{eqnarray*}
  \langle \underline{x}, \underline{p}, In(\underline{x}, \underline{p}), \{
  \tau_i(\underline{x}, \underline{p}, \underline{x}', \underline{p}')
  ~|~ i=1,...,n \}\rangle, 
\end{eqnarray*}
}%
where $\underline{x}$ is a tuple of \emph{WF state}
variables, $\underline{p}$ is a tuple of \emph{PM
state} variables, the \emph{initial condition} $In(\underline{x},
\underline{p})$ is a FOL formula whose only free variables are in
$\underline{x}$ and where PM state variables in $\underline{p}$ may
occur as predicate symbols, the \emph{transition} $\tau_i(\underline{x},
\underline{p}, \underline{x}', \underline{p}')$ is a FOL formula whose
only free variables are in $\underline{x},\underline{x}'$ and where 
PM state variables in $\underline{p},\underline{p}'$ may occur as
predicate symbols (as it is customary, unprimed variables in $\tau_i$
refer to the values of the state before the execution of the transition
while those primed to the values of the state afterward) for $i=1, ...,
n$ and $n\geq 1$.

We assume there exists a so-called first-order \emph{underlying
structure $\langle D, I \rangle$ of the transition system $Tr$}, where
$D$ is the domain of values and $I$ is the mapping from the signature to
functions and relations over $D$, and in which the state variables and
the symbols of the signature used to write the formulae $In$ and
$\tau_i$ for $i=1, ..., n$ are mapped. A \emph{state} of $Tr$ is a pair
$v:=(v_{\underline{x}}, v_{\underline{p}})$ of mappings:
$v_{\underline{x}}$ from the WF state variables to $D$ 
and $v_{\underline{p}}$ from the PM state variables to 
relations over $D$. 
A \emph{run} of $Tr$ is a (possibly infinite) sequence of states $v^0,
v^1, ..., v_n, ...$ such that (a) $v^0$ satisfies $In$, in symbols
$v^0\models In$, and (b) for every pair $v^i,v^{i+1}$ in the sequence,
there exists $j\in \{1, ..., n\}$ such that $v^i,v^{i+1}$ satisfies
$\tau_j$, in symbols $v^i,v^{i+1}\models \tau_j$, where the domain of
$v^i$ is $\underline{x}, \underline{p}$ and that of $v^{i+1}$ is
$\underline{x}',\underline{p}'$. Given a formula
$G(\underline{x},\underline{p})$, called the \emph{goal}, an instance
of the \emph{goal reachability problem} for $Tr$ consists of answering
the following question: does there exist a natural number $\ell\geq 0$
such that the formula

\vspace*{-0.4cm}
{\footnotesize
\begin{eqnarray}
  \label{eq:reachability-problem}
  In(\underline{x}_0,\underline{p}_0) \wedge 
  \bigwedge_{i=o}^{\ell-1} \tau(\underline{x}_i,\underline{p}_i,
                           \underline{x}_{i+1},\underline{p}_{i+1}) \wedge
  G(\underline{x}_{\ell},\underline{p}_{\ell})
\end{eqnarray}
}%
is satisfiable in the underlying structure of $Tr$, where
$\underline{x}_i,\underline{p}_i$ are renamed copies of the state
variables in $\underline{x},\underline{p}$? (When $\ell=0$, (\ref{eq:reachability-problem}) is simply $In(\underline{x}_0,\underline{p}_0) \wedge
G(\underline{x}_{0},\underline{p}_{0})$.) The interest of the goal
reachability problem lies in the fact that many 
verification problems for two-level transition systems, such as deadlock
freedom and invariant checking, can be reduced to it.

\subsection{Forward reachability and symbolic execution tree}
\label{subsec:fr-sst}

If we were able to check automatically the satisfiability of
(\ref{eq:reachability-problem}), an idea to solve the goal
reachability problem for two-level transition systems would be to
generate instances of (\ref{eq:reachability-problem}) for increasing
values of $\ell$.  However, this would only give us a semi-decision
procedure for the reachability problem. In fact, this method
terminates only when the goal is reachable from the initial state,
i.e.~when the instance of (\ref{eq:reachability-problem}) for a
certain value of $\ell$ is unsatisfiable in the underlying structure
of the transition system $Tr$. But, when the goal is not reachable,
the check will never detect the unsatisfiability and we will be bound
to generating an infinite sequence of instances of
(\ref{eq:reachability-problem}) for increasing values of $\ell$. That
is, the decidability of the satisfiability of
(\ref{eq:reachability-problem}) in the underlying structure of $Tr$ is
only a necessary condition for ensuring the decidability of the goal
reachability problem.

We can formalize this method as follows.  The \emph{post-image} of a
formula $K(\underline{x}, \underline{p})$ with respect to a transition
$\tau_i$ is

\vspace*{-0.4cm}
{\footnotesize
\begin{eqnarray*}
  Post(K,\tau_i) & := &
   \exists \underline{x}',\underline{p}'.(
     K(\underline{x}', \underline{p}') \wedge 
     \tau_i(\underline{x}',\underline{p}', \underline{x},\underline{p})) .
\end{eqnarray*}
}%
For the class of transition systems that we consider below, we are
always able to find 
FOL formulae that are equivalent to $Post(K,\tau_i)$.  Thus, the use
of the second-order quantifier over the predicate symbols in
$\underline{p}'$ should not worry the reader (see
\secref{subsec:rbac4bpel-tfr} for details).  Now, define the following
sequence of formulae by recursion: $FR^0(K,\tau) := K$ and
$FR^{i+1}(K,\tau) := Post^i(FR^i,\tau) \vee FR^{i}(K,\tau)$, for
$i\geq 0$ and $\tau:=\bigvee_{k=1}^n \tau_i$.  The formula
$FR^{\ell}(K,In)$ describes the set of states of the transition system
$Tr$ that are \emph{forward reachable in $\ell\geq 0$ steps}.  A
\emph{fix-point} is the least value of $\ell$ such that
$FR^{\ell+1}(\tau,In) \Rightarrow FR^{\ell}(\tau,In)$ is true in the
structure underlying $Tr$.  Note also that $FR^{\ell}(\tau,In)
\Rightarrow FR^{\ell+1}(\tau,In)$ by construction and hence if
$FR^{\ell+1}(\tau,In) \Rightarrow FR^{\ell}(\tau,In)$ is valid, then
also $FR^{\ell}(\tau,In) \Leftrightarrow FR^{\ell+1}(\tau,In)$ is so
and $FR^{\ell}(\tau,In) \Leftrightarrow FR^{\ell'}(\tau,In)$ for each
$\ell'\geq \ell$.  Using the sequence of formulae $FR^0(\tau,In),
FR^1(\tau,In), ...$ it is possible to check if the goal property $G$
will be reached by checking whether $FR^{\ell}(\tau,In)\wedge G$ is
satisfiable in the structure underlying $Tr$ for some $\ell\geq 0$.
In case of satisfiability, we say that $G$ is \emph{reachable}.
Otherwise, if $FR^{\ell}(\tau,In)$ is a fix-point, the
unsatisfiability of $FR^{\ell}(\tau,In)\wedge G$ implies that $G$ is
\emph{unreachable}.  Finally, if $FR^{\ell}(\tau,In)$ is not a
fix-point and $FR^{\ell}(\tau,In)\wedge G$ is unsatisfiable, then we
must increase the value of $\ell$ by $1$ so as to compute the set of
forward reachable states in $\ell+1$ steps and perform the
reachability checks again.  Unfortunately, also this process is not
guaranteed to terminate for arbitrary two-level transition systems.
Fortunately, we are able to characterize a set of transition systems,
corresponding to a relevant class of applications specified in BPEL
and RBAC4BPEL, for which we can pre-compute an upper bound on $\ell$;
this paves the way to solving automatically the goal reachability
problem for these systems.
To this end, we consider three sufficient conditions to automate the
solution of the goal reachability problem.  First, the class
$\mathcal{C}$ of formulae used to describe sets of states must be
closed under post-image computation.  Second, the satisfiability (in
the structure underlying the transition system) of $\mathcal{C}$ must
be decidable.  Third, it must be possible to pre-compute a bound on
the length of the sequence $FR^0, FR^1, ..., FR^{\ell}$ of formulae.
Below, we show that these conditions are satisfied by a class of
two-level transition systems to which applications specified in BPEL
and RBAC4BPEL can be mapped.  For ease of exposition, we first
consider the WF and PM levels in isolation and then show how the
results for each level can be modularly lifted when considering the
two levels together.  Before doing this, we introduce the notion of
`symbolic execution tree.'  The purpose of this is two-fold.  First,
it is crucial for the technical development of our decidability
result.  Second, it is the starting point for the implementation of
our techniques as discussed in~\secref{sec:industrial}.

The \emph{symbolic execution tree of the two-level transition system
  $Tr$} is a labeled tree defined as follows: (i) the root node is
labeled by the formula $In$, (ii) a node $n$ labeled by the formula
$K$ has $d\leq n$ sons $n_1, ..., n_d$ labeled by the formulae
$Post(\tau_1, K), ..., Post(\tau_d, K)$ such that $Post(\tau_j, K)$ is
satisfiable in the model underlying $Tr$ and the edge from $n$ to
$n_j$ is labeled by $\tau_j$ for $j=1, ..., d$, (iii) a node $n$
labeled by $K$ has no son, in which case $n$ is a \emph{final node},
if $Post(\tau_j, K)$ is unsatisfiable in the underlying model of the
VAS, for each $j=1, ..., n$. A symbolic execution tree is
\emph{$0$-complete} if it consists of the root node labeled by the
formula $In$, it is \emph{$(d+1)$-complete} for $d\geq 0$ if its depth
is $d+1$ and for each node $n$ labeled by a formula $K_n$ at depth
$d$, if $Post(\tau_j,K_n)$ is satisfiable, then there exists a node
$n'$ at depth $d+1$ labeled by $Post(\tau_j,K_n)$.  In other words, a
symbolic execution tree is $d$-complete when all non-empty sets of
forward states reachable in one step represented by formulae labeling
nodes at depth $d-1$ have been generated. It is easy to see that the
formula $FR^{\ell}(K,In)$, describing the set of states of the
transition system $Tr$ {forward reachable in $\ell\geq 0$ steps, is
  equivalent to the disjunction of the formulae labeling the nodes of
  an $\ell$-complete symbolic execution tree}. This will be proved for
the classes of two-level transition systems that we consider below.

\subsection{WF nets and terminating forward reachability}
\label{subsec:wn-tfr}

We consider a particular class of two-level transition systems, called
\emph{Vector Addition System (VAS)}, $\langle \underline{x},
In(\underline{x}), \{ \tau_i(\underline{x}, \underline{x}') ~|~
i=1,...,n \}\rangle$, such that (a) $\underline{p}=\emptyset$; (b) their
underlying structure is that of integers; (c) each WF state variable in
$\underline{x} = x_1, ..., x_m$ ranges over the set of non-negative
integers; (d) the initial condition $In(\underline{x})$ is a formula of
the form $x_i \bowtie c_1 \wedge \cdots \wedge x_m \bowtie c_m$, where
$c_j$ is a natural number for $j=1, ..., m$ and $\bowtie\in \{ =, \neq, >,
\geq \}$; and (e) each transition $\tau_i$, for $i=1,...,n$, is a
formula of the form

\vspace*{-0.4cm}
{\footnotesize
\begin{eqnarray*}
  \bigwedge_{i\in P} x_i\geq 0 \wedge 
  \bigwedge_{j\in U^+} x_j'=x_j+1 \wedge 
  \bigwedge_{k\in U^-} x_k'=x_k-1 \wedge
  \bigwedge_{l\in U^=} x_l'=x_l ,
\end{eqnarray*}
}%
where $P, U^+,U^-,U^=$ are subsets of $\{1, ..., n\}$ such that
$U^+,U^-,U^=$ form a partition of $\{1, ..., n\}$.  

It is well-known that Petri nets and VASs are equivalent in the sense
that analysis problems for the former can be transformed to problems
of the latter whose solutions can be mapped back to solutions for the
original problem and vice versa (see,
e.g.,~\cite{manna-petri-net-invariant-generation}).  We briefly
describe the correspondence by considering the Petri net in
Fig.~\ref{fig:ex-forms-pn}.  
We associate an integer variable $x_i$ to each place $p_i$ for
$i=1,...,10$ whose value will be the number of tokens in the place.
The state is given by the value of the integer variables that
represents the marking of the net, i.e.\ a mapping from the set of
places to non-negative integers.  Formulae can be used to represent
sets of states (or, equivalently, of markings). So, for example, the
formula $x_1=1\wedge \bigwedge_{i=2}^{10} x_i=0$ represents the
marking where one token is in place $p_1$ and all the other places are
empty (which is the one depicted in Fig.~\ref{fig:ex-forms-pn} where
the token is represented by a solid circle inside that represents
the place $p_1$ while all the other places do not contain any solid
circle).  The transition crtPO is represented by the formula

\vspace*{-0.4cm}
{\footnotesize
\begin{eqnarray*}
  x_1\geq 1 \wedge 
  x_1'=x_1-1\wedge x_2'=x_2+1\wedge \bigwedge_{i=3}^{10}  x_i=x_i
\end{eqnarray*}
}%
saying that it is enabled when there is at least one token in $p_1$
($x_1\geq 1$) and the result of its execution is that a token is
consumed at place $p_1$ ($x_1'=x_1-1$), the tokens in $p_2$ are
incremented by one ($x_2'=x_2+1$), while the tokens in all the other
places are unaffected ($x_i'=x_i$ for $i=3, ..., 10$). The other
transitions of the Petri net in Fig.~\ref{fig:ex-forms-pn} are translated
in a similar way. In
general, it is always possible to associate a state of a VAS to a
marking of a Petri net and vice versa. This implies that solving the
reachability problem for a VAS is equivalent to solving the reachability
problem of the associated Petri net.

Now, we show that the three sufficient conditions (see \secref{subsec:fr-sst}) to mechanize the
solution of the goal reachability problem 
are satisfied by VASs when using forward reachability.  First, the
class of formulae is closed under post-image computation.
\begin{fact}
  \label{fact:post-VAS}
  $Post(K,\tau_i)$ is equivalent to $K[x_{j}+1, {x}_{k}-1, {x}_{l}]
  \wedge \bigwedge_{i\in P} x_i\geq 0$, where $K[x_{j}+1, {x}_{k}-1,
  {x}_{l}]$ denotes the formula obtained by replacing $x_j'$ with
  $x_{j}-1$ for $j\in U^+$, $x_k'$ with $x_{k}-1$ for $k\in U^-$, and
  $x_l'$ with $x_{l}+1$ for $j\in U^=$. \hfill $\blacksquare$
\end{fact}

As a corollary, it is immediate to derive that if $K$ is a formula of
Linear Arithmetic (LA)~\cite{linear-arithmetic}---roughly, a formula
where multiplication between variables is forbidden---then also
$Post(K,\tau_i)$ is equivalent to an effectively computable formula of
LA. Second, the satisfiability of the class of formulae of LA is
decidable by well-known results~\cite{linear-arithmetic}.
Third, it is possible to pre-compute a bound on the length of the
sequence $FR^0, FR^1, ..., FR^{\ell}$ of formulae. Using the notion of
symbolic execution tree introduced above, once specialized to VASs, we
can then prove:
\begin{lemma}
  \label{lem:acyclic-workflow-net}
  Let $PN:=\langle P, T, F \rangle$ be an acyclic workflow net and
  $\Pi$ be the set of all its paths.  Then, the set of forward
  reachable states of the VAS $\langle \underline{x},
  In(\underline{x}), \{ \tau_i(\underline{x}, \underline{x}') ~|~
  i=1,...,n \}\rangle$ associated to $PN$ is identified by the formula
  $FR^{\ell}(\tau, In)$ for $\ell = max_{\pi\in\Pi} \{ len(\pi|_{T})
  \}$, where $\pi|_{T}$ is the sequence obtained from $\pi$ by
  forgetting each of its elements in $P$ and $len(\pi|_T)$ is the
  length of the sequence $\pi|_T$.  \hfill $\blacksquare$
\end{lemma}

\subsection{RBAC4BPEL and terminating forward reachability}
\label{subsec:rbac4bpel-tfr}

Preliminarily, let $Enum(\{v_1, ..., v_n \}, S)$ be the following set
of FOL formulae axiomatizing the enumerated datatype with values $v_1,
..., v_n$ for a given $n\geq 1$ over a type $S$: $v_i\neq v_j$ for
each pair $(i,j)$ of numbers in $\{1, ..., n\}$ such that $i\neq j$
and $\forall x.\, (x=v_1 \vee \cdots\vee x=v_n)$, where $x$ is a
variable of type $S$.  The formulae in $Enum(\{v_1, ..., v_n \}, S)$
fix the number of elements of any interpretation to be $v_1, ...,
v_n$; it is easy to see that the class of structures satisfying these
formulae are closed under isomorphism.  We consider a particular class
of two-level transition systems, called RBAC4BPEL, $\langle
\underline{p}, In(\underline{p}), \{ \tau_i(\underline{p},
\underline{p}') ~|~ i=1,...,n \}\rangle$ such that (a)
$\underline{x}=\emptyset$; (b) the initial condition
$In(\underline{p})$ is of the form $\forall \underline{w}.\,
\varphi(\underline{w})$, where $\varphi$ is a quantifier-free formula
where at most the variables in $\underline{w}$ may occur free; and (c)
the underlying structure is one in the (isomorphic) class of
many-sorted structures axiomatized by the following sentences:

\vspace*{-0.4cm}
{\footnotesize
\begin{eqnarray*}
  Enum(U, \mathit{User}), ~~
  Enum(R, \mathit{Role}), \\
  Enum(P, \mathit{Permission}), ~~
  Enum(A, \mathit{Action}),  \\
  \forall u,r.(ua(u,r) \Leftrightarrow 
    \bigvee_{c_U\in U_{ua}, c_r\in R_{ua}} (u=c_u \wedge r=c_r)) \\
  \forall r,p.(pa(r,p) \Leftrightarrow 
    \bigvee_{c_r\in R_{pa}, c_p\in P_{pa}} (r=c_r \wedge p=c_p)) \\
  c_r\succeq c_r' \mbox{ for }c_r,c_r'\in R  \\
  \forall r.(r\succeq r) ~~
  \forall r_1,r_2,r_3.(r_1\succeq r_2\wedge r_2\succeq r_3\Rightarrow r_1\succeq r_3)\\
  \forall r_1,r_2.(r_1\succeq r_2\wedge r_2\succeq r_1\Rightarrow r_1=r_2),
\end{eqnarray*}
}%
where $U$, $R$ and $P$ are finite sets of constants denoting users,
roles, and permissions, respectively, $A$ is a finite set of actions,
$u$ is a variable of type $\mathit{User}$, $r$ and its subscripted
versions are variables of type $\mathit{Role}$, $p$ is a variable of
type $\mathit{Permission}$, $U_{ua}\subseteq U$, $R_{ua}\subseteq R$,
$R_{pa}\subseteq R$, and $P_{pa}\subseteq P$; (d)
$\underline{p}=\mathit{xcd}$ is a predicate symbol of type
$\mathit{User}\times \mathit{Action}$ abbreviating
$\mathit{executed}$; and (e) each $\tau_i$ is of the form

\vspace*{-0.4cm}
{\footnotesize
\begin{eqnarray*}
  \exists \underline{u}.\, (\xi(\underline{u},\mathit{xcd}) \wedge 
             \forall x,y.(\mathit{xcd}'(x,y) \Leftrightarrow 
                          ((x=u_j\wedge y=p) \vee \mathit{xcd}(x,y)))), 
\end{eqnarray*}
}%
where $\underline{u}$ is a tuple of existentially quantified variables
of type $\mathit{User}$, $u_j$ is the variable at position $j$ in
$\underline{u}$, and $\xi(\underline{u})$ is a quantifier-free formula
(called the \emph{guard} of the transition) where no function symbol
of arity greater than $0$ may occur (the part of $\tau_i$ specifying
$\mathit{xcd}'$ is called the \emph{update}).

We now explain how an RBAC4BPEL system can be specified by the formulae
above on the example described in \secref{sec:motivating}. To constrain
the sets of users, of roles, and permissions to contain exactly the
elements specified in Fig.~\ref{fig:ex-forms-rbac}, it is
sufficient to use the following sets of formulae: $Enum(\{u_1, u_2, u_3,
u_4\}, \mathit{User})$, $Enum(\{Manager, FinAdmin, FinClerk,
POAdmin,PO$- $Clerk\},$ $\mathit{Role})$, and $Enum(\{p_1, p_2, p_3,
p_4, p_5\}, \mathit{Permission})$. It is also easy to see that the
formulae

\vspace*{-0.4cm}
{\footnotesize
\begin{eqnarray*}
  \forall u,r.(ua(u,r) \Leftrightarrow 
   \left(
   \begin{array}{l}
     (u=u_1\wedge r=Manager) \vee \\
     (u=u_2\wedge r=FinAdmin)\vee \\
     (u=u_3\wedge r=FinClerk)\vee \\
     (u=u_4\wedge r=POAdmin)\vee \\
     (u=u_5\wedge r=POClerk)
   \end{array}
   \right) \\
  \forall r,p.(pa(r,p) \Leftrightarrow 
   \left(
   \begin{array}{l}
     (r=FinClerk\wedge p=p_4)\vee\\
     (r=FinAdmin\wedge p=p_5)\vee\\
     (r=POClerk\wedge  p=p_3)\vee\\
     (r=POAdmin\wedge  p=p_1)
   \end{array}
   \right) \\
\end{eqnarray*}
}%
are satisfied by the interpretations of $ua$ and $pa$ in
Fig.~\ref{fig:ex-forms-rbac} 
and that $Manager\succeq FinAdmin$, $Manager\succeq POAdmin$,
$FinAdmin\succeq FinClerk$, and $POAdmin\succeq POClerk$ with the
three formulae above for reflexivity, transitivity and antisymmetry
make the interpretation of $\succeq$ the partial order considered in
Fig.~\ref{fig:ex-forms-rbac}.  The state variable $\mathit{xcd}$ allows us
to formalize component (vii) of the RBAC4BPEL system about the
authorization constraints.  The idea is to use $\mathit{xcd}$ to store
the pair user $u$ and action $a$ when $u$ has performed $a$ so that
the authorization constraints can be formally expressed by a
transition involving suitable pre-conditions on these variables.  We
illustrate the details on the first authorization constraint
considered in \secref{subsec:rbac4bpel}, i.e.\ $\langle U,
(apprPO,signGNR), \neq \rangle$.  The corresponding transition can be
formalized as follows:

\vspace*{-0.4cm}
{\footnotesize
\begin{eqnarray*}
  \exists x_1,x_2.(\mathit{xcd}(x_1,apprPO)\wedge x_1\neq x_2 \wedge \\
   \forall x,y.(\mathit{xcd}'(x,y) \Leftrightarrow ((x=x_2\wedge y=signGNR)\vee \mathit{xcd}(x,y))) .
\end{eqnarray*} 
}%
The guard of the transition prescribes that the user $x_2$ is not the
same user $x_1$ that has previously performed the action apprPO and
the update stores in $\mathit{xcd}$ the new pair $(x_2,signGNR)$.  The
following two constraints at the end of \secref{subsec:rbac4bpel}, namely
$\langle U, (apprPO,ctrSignGNR), \neq \rangle$ and $\langle U,
(signGNR,ctrSignGNR), \neq \rangle$, are formalized in a similar way.
The encoding of the last constraint, i.e.\ $\langle R, (ctrPay,
apprPay), \prec \rangle$, is more complex and requires also the use of
the user-role relation $ua$ to represent the constraint on the role
hierarchy:

\vspace*{-0.4cm}
{\footnotesize
\begin{eqnarray*}
  \exists x_1,x_2,r_1,r_2.( \mathit{xcd}(x_1,crtPay)\wedge ua(x_1,r_1)\wedge\\
  ua(x_2,r_2)\wedge r_2\succeq r_1\wedge r_1\neq r_2 \wedge \\
   \forall x,y.(\mathit{xcd}'(x,y) \Leftrightarrow ((x=x_2\wedge y=apprPay)\vee \mathit{xcd}(x,y))) .
\end{eqnarray*} 
}%
The reader should now be convinced that 
every RBAC4BPEL specification can be translated into a RBAC4BPEL system.

Now, we show that the three sufficient conditions to mechanize the
solution of the goal reachability problem (see \secref{subsec:fr-sst})
are satisfied by RBAC4BPEL systems when using forward reachability.
First, the class of formulae is closed under post-image computation.
\begin{fact}
  \label{fact:post-RBAC}
  $Post(K,\tau_i)$ is equivalent to 

\vspace*{-0.4cm}
{\footnotesize  \begin{eqnarray*}
    (\exists \underline{u}.(K(\mathit{xcd}) \wedge
    \mathit{xcd}({u}_j,t) \wedge
    \xi(\underline{u},\mathit{xcd}) )) & \vee \\
    (\exists \underline{u}.(K[\lambda x,y.(\neg (x=u_j\wedge y=t) \wedge 
    \mathit{xcd}(x,y))] \wedge \\ 
    \xi[\underline{u},\lambda x,y.(\neg (x=u_j\wedge y=t) \wedge 
    \mathit{xcd}(x,y))])) & , 
  \end{eqnarray*}
}%
  where $K[\lambda x,y.(\neg (x=u_j\wedge y=t) \wedge
  \mathit{xcd}(x,y))]$ is the formula obtained from $K$ by
  substituting each occurrence of $xcd'$ with the $\lambda$-expression
  in the square brackets and then performing the $\beta$-reduction and
  similarly for $\xi[\underline{u},\lambda x,y.(\neg (x=u_j\wedge y=t)
  \wedge \mathit{xcd}(x,y))]$. \hfill $\blacksquare$
\end{fact}

As anticipated above when introducing the definition of post-image for
two-level transition systems, we can eliminate the second-order
quantifier over the predicate symbol $\mathit{xcd}$. Now, recall that a
formula is in the \emph{Bernays-Sch\"onfinkel-Ramsey (BSR)} class if it
has the form $\exists \underline{z}\forall\underline{w}.\,
\phi(\underline{z},\underline{w})$, for $\phi$ a quantifier-free formula
and $\underline{z}\cap\underline{w}=\emptyset$ (see, e.g.,~\cite{epr}).
As a corollary of Fact~\ref{fact:post-RBAC}, it is immediate to see
that if $K$ is a BSR formula, then also $Post(\tau_i,K)$ is
equivalent---by trivial logical manipulations---to a formula in the BSR
class. Since $In(\mathit{xcd})$ is a formula in the BSR class, then all
the formulae in the sequence $FR^0, FR^1, ...$ will also be BSR
formulae. The second requirement is also fulfilled since the
satisfiability  
of the BSR class is well-known to be decidable~\cite{epr} and the
formulae used to axiomatize the structures underlying the RBAC4BPEL
transition systems are also in BSR. Third, it is possible to
pre-compute a bound on the length of the sequence $FR^0, FR^1, ...,
FR^{\ell}$ of formulae, although the existential prefix grows after
each computation of the post-image when considering the formulae
describing the set of forward reachable states. This is so because we
consider only a finite and known set of users so that the length of
the existentially quantified prefix is bounded by $n_u^k\times n$,
where $k$ is the maximal length of the existential prefixes of the
transitions in the RBAC4BPEL system, $n_u$ is the number of users, and
$n$ is the number of transitions.
\begin{property}
  \label{prop:RBAC-tree-completeness}
  Let $\langle \underline{p}, In(\underline{p}), \{
  \tau_i(\underline{p}, \underline{p}') ~|~ i=1,...,n \}\rangle$ be a
  RBAC4BPEL system, $k$ the maximal length of the existential prefixes
  of $\tau_1, ..., \tau_n$, and $n_u$ be the cardinality of the set of
  users.  Then, its symbolic execution tree is $\ell$-complete for
  every $\ell\geq n_u^k\times n$. \hfill $\blacksquare$
\end{property}

The key idea of the proof is the observation that $\mathit{xcd}$ is
interpreted as a subset of the Cartesian product between the set of
users and the set of actions whose cardinalities are bounded.

\subsection{Combining VASs and RBAC4BPEL systems}
\label{subsec:combining-vas-RBAC4BEPL}

We are now ready to fully specify applications that feature both the
WF and the PM level.  To do this, we consider two-level transition
systems, called \emph{VAS+RBAC4BPEL} systems, of the form

\vspace*{-0.4cm}
{\footnotesize
\begin{eqnarray*}
  \langle \underline{x}, \underline{p},
  In_{V}(\underline{x})\wedge In_{R}(\underline{p}),
  \{ \tau_i^{V}(\underline{x},\underline{x}') \wedge 
     \tau_i^{R}(\underline{p},\underline{p}') ~|~ i=1,...,n \}\rangle ,
\end{eqnarray*}
}%
where $\underline{x}=x_1, ..., x_n$ for some $n\geq 1$,
$\underline{p}=\mathit{xcd}$, $In_{V}(\underline{x})$ is the initial
condition of a VAS, $In_{R}(\underline{p})$ is the initial condition
of a RBAC4BPEL system, $\tau_i^{V}(\underline{x},\underline{x}')$ is a
transition of a VAS, $\tau_i^{R}(\underline{p},\underline{p}')$ is a
transition formula of a RBAC4BPEL system for $i=1, ..., n$.  Note that
for some transition, the guard $\xi$ of
$\tau_i^{R}(\underline{p},\underline{p}')$ may be tautological since
the operation involves no access-control policy restriction (e.g., the
`flow split' and `flow join' of the Petri net in
Fig.~\ref{fig:ex-forms-pn}). It is natural to associate a VAS and an
RBAC4BPEL system to a VAS+RBAC4BPEL system by projection, i.e.\ the
associated VAS is $ \langle \underline{x}, In_{V}(\underline{x}), \{
\tau_i^{V}(\underline{x},\underline{x}') ~|~ i=1,...,n \}\rangle$ and
the associated RBAC4BPEL system is $ \langle \underline{p},
In_{R}(\underline{p}), \{ \tau_i^{R}(\underline{p},\underline{p}') ~|~
i=1,...,n \}\rangle$. The structure underlying the VAS+RBAC4BPEL
system is such that its reduct to the signature of the VAS is
identical to the structure underlying the associated VAS and its
reduct to the signature of the RBAC4BPEL system is identical to the
structure underlying the associated RBAC4BPEL system.

We now show how it is possible to modularly compute the post-image of
a VAS+RBAC4BPEL system by combining the post-images of the associated
VAS and RBAC4BPEL system.  
\begin{fact}
  \label{fact:modular-post-image}
  Let $K(\underline{x},\mathit{xcd}) := K_{V}(\underline{x})\wedge
  K_{R}(\mathit{xcd})$.  Then, $Post(K,\tau_i)$ is equivalent to

\vspace*{-0.4cm}
{\footnotesize
  \begin{eqnarray*}
    K_{V}[x_{j}+1, {x}_{k}-1, {x}_{l}]
    \wedge \bigwedge_{i\in P} x_i\geq 0 & \wedge \\
    ((\exists \underline{u}.(K_{R}(\mathit{xcd}) \wedge
    \mathit{xcd}({u}_j,t) \wedge
    \xi(\underline{u},\mathit{xcd}) )) & \vee \\
    (\exists \underline{u}.(K_{R}[\lambda x,y.(\neg (x=u_j\wedge y=t) \wedge 
    \mathit{xcd}(x,y))] \wedge \\ 
    \xi[\underline{u},\lambda x,y.(\neg (x=u_j\wedge y=t) \wedge 
    \mathit{xcd}(x,y))]))) & , 
  \end{eqnarray*}
}%
where the same notational conventions of Facts~\ref{fact:post-VAS}
and~\ref{fact:post-RBAC} have been adopted.  In other words, the
post-image of a VAS+RBAC4BPEL system is obtained as the conjunction of
the post-images of the associated VAS, denoted with
$Post_V(K,\tau_i):=Post(K_V,\tau_i^V)$, and the associated RBAC4BPEL
system, denoted with $Post_R(K,\tau_i):=Post(K_R,\tau_i^R)$.  Thus, we
abbreviate the above formula as $Post_V(K,\tau_i)\wedge
Post_R(K,\tau_i)$. \hfill $\blacksquare$
\end{fact}

The proof of this fact is obtained by simple manipulations minimizing
the scope of applicability of $\exists \underline{x}$ and $\exists
\mathit{xcd}$, respectively, and then realizing that the proofs of
Facts~\ref{fact:post-VAS} and~\ref{fact:post-RBAC} can be re-used
verbatim. Because of the modularity of post-image computation, it is
possible to modularly define the set of forward reachable states and the
symbolic execution trees for VAS+RBAC4BPEL systems in the obvious way.
By modularity, 
we can easily show the following property.
\begin{property}
  \label{prop:finiteness-symb-sim-tree-VAS+RBAC}
  Let $PN:=\langle P,T,F \rangle$ be a an acyclic WF net, $\langle
  \underline{x}, In_V(\underline{x}), \{
  \tau_i^V(\underline{x},\underline{x}') ~|~ i=1, ..., n\} \rangle$ be
  its associated VAS, and $\langle \underline{p}, In_R(\underline{p}),
  \{ \tau_i^R(\underline{p},\underline{p}') ~|~ i=1, ..., n\} \rangle$
  be the RBAC4BPEL system with $n_u$ users and $k$ be the maximal
  length of the existential prefixes of $\tau_1^R, ..., \tau_n^R$.
  Then, the symbolic reachability tree of the VAS+RBAC4BPEL system
  whose associated VAS and RBAC system are those specified above is
  $\ell$-complete for every $\ell\geq min(max_{\pi\in\Pi} \{
  len(\pi|_{T}) \}, n_u^k\times |T|)$. \hfill
  $\blacksquare$
\end{property}

The key observation in the proof of this property is that in order to
take a transition, the preconditions of the associated VAS and of the
associated RBAC4BPEL system must be satisfied. Because of the modularity
of the post-image, the duality between the set of forward reachable
states and the formulae labeling the symbolic execution tree can be
lifted to VAS+RBAC4BPEL.
We are now ready to state and prove the main result of this paper.
\begin{theorem}
  \label{th:main}
  Let $PN:=\langle P,T,F \rangle$ be a an acyclic WF net and let
  $\langle \underline{x}, In_V(\underline{x}), \{
  \tau_i^V(\underline{x},\underline{x}') ~|~ i=1, ..., n\} \rangle$ be
  its associated VAS.  Further, let $\langle \underline{p},
  In_R(\underline{p}), \{ \tau_i^R(\underline{p},\underline{p}') ~|~
  i=1, ..., n\} \rangle$ be an RBAC4BPEL system with a bounded number
  of users.  Then, the symbolic reachability problem of the
  VAS+RBAC4BPEL system (whose associated VAS and RBAC4BPEL system are
  those specified above) is decidable.  \hfill $\blacksquare$
\end{theorem}

To mechanize this result, we can use off-the-shelf a
state-of-the-art Satisfiability Modulo Theories solver such as Z3~\cite{Z3}
that are capable of automatically discharging the proof obligation
generated by the iterated computation of the post-image in the
structures underlying the VAS+RBAC4BPEL system.

To illustrate the kind of formulae arising in the application of
Theorem~\ref{th:main}, we consider the example specified in
Fig.~\ref{fig:ex-forms-pn}. In this case, we can restrict to consider
three paths (projected over the transitions) in the WF net: crtPO,
apprPO, `flow split', signGRN, ctrSignGNR, crtPay, `flow join',
apprPay; crtPO, apprPO, `flow split', signGRN, crtPay, crtSignGRN,
`flow join', apprPay; and crtPO, apprPO, `flow split', crtPay,
signGRN, ctrsignGRN, `flow join', apprPay; each one of length
eight. It is easy to see that only the first path is to be considered
as the other two produce states that are equivalent since it does not
matter at what time crtPay is executed with respect to signGRN and
ctrSignGNR (it is possible to mechanize also this check but we leave
out the details for lack of space). So, for example, it is possible to
check the so-called soundness of workflows~\cite{mc-ws}, i.e.\ to
check whether it is possible to terminate without ``garbage'' left.
In terms of a WF net, this means that no tokens are left in places
other than the special final place $o$ of the net. This can be checked
by computing the post-images of the initial state of the VAS+RBAC4BPEL
system of our motivating example along the lines of
Facts~\ref{fact:post-VAS},~\ref{fact:post-RBAC},
and~\ref{fact:modular-post-image} and put this in conjunction with the
formula characterizing the ``no-garbage'' condition, i.e.\

\vspace*{-0.4cm}
{\footnotesize
\begin{eqnarray*}
  x_{10}\geq 1\wedge \bigwedge_{i=1}^{9} x_i=0 .
\end{eqnarray*}
}%
The resulting proof obligation, because of the closure under post-image
computation of the VAS and the RBAC4BPEL system as well as the
modularity of the post-image computation for the VAS+RBAC4BPEL system,
is decidable as it can be put in the form $\varphi_{V}\wedge
\varphi_{R}$ where $\varphi_V$ is a formula of LA (whose satisfiability
is decidable) and $\varphi_R$ is a BSR formula (whose satisfiability is
again decidable), and thus the satisfiability of their conjunction is
also decidable.

\section{Analysis of an industrial case study}
\label{sec:industrial}

We have implemented a prototype tool, called \emph{WSSMT}, that allows
the user to explore the symbolic execution tree of a VAS+RBAC4BPEL
system. WSSMT features a client-server architecture where the server is
the Z3 SMT solver while the client (implemented in Java as an Eclipse
plug-in) takes a two-level transition system and generates the proof
obligations for solving the reachability problem as discussed in
Theorem~\ref{th:main}.

We have first applied WSSMT on the example described in
\secref{sec:motivating} to validate our ideas and then we have
considered a more significant example, inspired by an industrial case
study, i.e.\ the Digital Contract Signing (DCS, for short).  The
scenario consists of two signers having secure access to a trusted
third party, called a Business Portal (BP), in order to digitally sign
a contract.
To achieve this goal, 
each signer sets the contract's conditions by communicating them to
BP, which creates a digital version of the contract, stores
it, 
and coordinates the two signers in order to obtain their signatures.
The DCS process is successful when both signers provide genuine
signatures for the digital contract and the BP can permanently store
the signed copy of the contract.


The WF level specification of the DCS consists of four BPEL processes:
one for the BP, one for the two instances of the signers, one for the
service checking the signature, and one for the service archiving the
contract. To create the composed BPEL process out of the four
components, we have used the BPEL2oWFN tool~\cite{bpel2owfn} that is
also capable of generating a Petri net representation of the resulting
process. We have modified the tool in order to generate the associated
VAS as described in \secref{subsec:wn-tfr}.
As a result, we have obtained a VAS with $50$ integer variables and
$26$ transitions.

The PM level specification of the DCS has been manually specified as
there seem to be no available tool for mechanizing this task.  More
precisely, we have specified an RBAC4BPEL system along the lines of
\secref{subsec:rbac4bpel-tfr}.  The set $U$ of users is composed of
five users: two signers, the BP, one checking the signature, and one
archiving the contract; the set $R$ of roles contains four roles
corresponding to each BPEL process; the set $P$ of permissions lists
$24$ elements corresponding to the right of executing the $26$
transitions ($2$ transitions do not need authorization constraints
because they are `flow split' and `flow join' as in the Petri net in
Fig.~\ref{fig:ex-forms-pn} and are thus used only for synchronization at
the WF level); the relation $ua$ prescribes the obvious associations
between users and roles (e.g., the two users willing to sign the
contract belong to the role of signers); and the relation $pa$ also
associates the $24$ permissions to the $24$ transitions that need
authorization constraints.
Finally, we have added SoD (e.g., the user signing the contract should
not be the same as the one checking the validity of the signature on
the contract) and BoD (e.g., the users signing the contract should be
same that have agreed on the conditions of the contract) authorization
constraints.

The property that we would like to check for the DCS is that once a
signed contract has been permanently stored, its signatures have been
checked valid and belong to the users who provided the conditions in
the contract. Indeed, to be formalized and then verified, this
property requires the specification of the manipulation on the data
(mainly, the contract) exchanged by the various BPEL processes. As we
already observed, this is difficult if not impossible for tools like
BPEL2oWFN as they consider only the control flow. One of the main
advantages of using (fragments of) FOL as done in this paper is the
flexibility of adding features to an available specification so as to
refine it and to allow for the verification of more complex properties
such as the one mentioned above. As a consequence, we have manually
added to the available specification of the DCS a description of the
messages exchanged among the various processes and how they are
generated or modified by using well-known techniques for the
specification of message-passing systems in FOL (see,
e.g.,~\cite{uclid}). For example, we were able to characterize the
BPEL notion of `correlation set', i.e.\ messages passed around contain
key fields (e.g., user IDs or any business-application-specific
identifiers) that can be correlated for the lifetime of the exchange
and, e.g., enabled the BP to distinguish the messages sent by the
first signer from those of the second. This kind of information, which
is crucial for proving properties of the kind specified above, is lost
while generating the Petri net by using tools such as BPEL2oWFN.

As a result, with our refined model of the DCS, we were able to verify
the property about the contracts that are permanently stored given above
in less than $10$ seconds on a standard laptop. This is an encouraging
result about the scalability of our techniques.

\section{Conclusion}
\label{sec:conclusion}

We have described automated analysis techniques for the validation of a
class of web services specified in BPEL and RBAC4BPEL. We have used
decidable fragments of FOL to describe the state space of this class of
services and then used the state-of-the-art SMT solver Z3 to solve their
reachability problems. We have applied our techniques to the
verification of a digital contract signing service by using a prototype
tool. The success in solving this verification is due to the flexibility
of our specification framework that allowed us to precisely capture the
interplay between the control flow, the data flow, and the
access-control level of the service. As future work, we plan, for
instance, to extend our decidability results to WF nets containing
restricted form of loops and extensions of RBAC4BPEL with delegation.

\section*{Acknowledgment}

The work presented in this paper was partially supported by the
FP7-ICT-2007-1 Project no.~216471, ``AVANTSSAR: Automated Validation of
Trust and Security of Service-oriented Architectures'' and the
``Automated Security Analysis of Identity and Access Management Systems
(SIAM)'' project funded by Provincia Autonoma di Trento in the context
of the ``Team 2009 - Incoming'' COFUND action of the European Commission
(FP7). We thank Luca Zanetti for the implementation of the WSSMT tool.





%

\begin{SHORT}
\bibliographystyle{IEEEtran}
\bibliography{biblio}
\end{SHORT}

\begin{LONG}


\appendix 

\section*{Proofs of \secref{subsec:wn-tfr}}

\paragraph*{Fact~\ref{fact:post-VAS}}  
$Post(K,\tau_i)$ is equivalent to $K[x_{j}+1, {x}_{k}-1, {x}_{l}]
\wedge \bigwedge_{i\in P} x_i\geq 0$, where $K[x_{j}+1, {x}_{k}-1,
{x}_{l}]$ denotes the formula obtained by replacing $x_j'$ with
$x_{j}-1$ for $j\in U^+$, $x_k'$ with $x_{k}-1$ for $k\in U^-$, and
$x_l'$ with $x_{l}+1$ for $j\in U^=$.
\begin{IEEEproof} 
  Preliminarily, note that if $x_j=x_j'+1$ then $x_j'=x_j-1$ for
  $j\in U^+$, and if $x_k=x_k'-1$ then $x_k'=x_k+1$ for $k\in U^-$.
  Then, observe the following simple calculations:

\vspace*{-0.4cm}
{\footnotesize
  \begin{eqnarray*}
   &  & Post(K,\tau_i) \\
   & [\mbox{by definition of }Post]~~ \Leftrightarrow  &
   \exists x_1', ..., x_n'.(K(\underline{x}') \wedge 
                           \tau_i(\underline{x}',\underline{x})) \\
  & [\mbox{by definition of }\tau_i]~~ \Leftrightarrow  &
  \exists x_1', ..., x_n'.(K(x_1', ..., x_n') \wedge \\
  && \bigwedge_{i\in P} x_i\geq 0 \wedge  
  \bigwedge_{j\in U^+} x_j=x_j'+1 \wedge \\
  && \bigwedge_{k\in U^-} x_k=x_k'-1 \wedge 
  \bigwedge_{l\in U^=} x_l=x_l' ) \\
  & [\mbox{by replacement}]~~ \Leftrightarrow &  
  \exists \underline{x}'.(K[x_{j}+1, {x}_{k}-1, {x}_{l}] \wedge \\
  &&
  \bigwedge_{i\in P} x_i\geq 0 \wedge  
  \bigwedge_{j\in U^+} x_j=x_j'+1 \wedge \\
  && \bigwedge_{k\in U^-} x_k=x_k'-1 \wedge 
   \bigwedge_{l\in U^=} x_l=x_l' ) \\  
  & [\mbox{by a property of }\exists]~~ \Leftrightarrow &  
  K[x_{j}+1, {x}_{k}-1, {x}_{l}] \wedge 
  \bigwedge_{i\in P} x_i\geq 0 \wedge  \\
  &&
  \exists \underline{x}'.(\bigwedge_{j\in U^+} x_j=x_j'-1 \wedge \\
  &&\bigwedge_{k\in U^-} x_k=x_k'+1 \wedge 
   \bigwedge_{l\in U^=} x_l=x_l' ) \\  
  & [\mbox{by a property of }\exists]~~ \Leftrightarrow &  
  K[x_{j}+1, {x}_{k}-1, {x}_{l}] \wedge 
  \bigwedge_{i\in P} x_i\geq 0 .
\end{eqnarray*}
}%
This concludes the proof.
\end{IEEEproof}

\paragraph*{Lemma~\ref{lem:acyclic-workflow-net}}
Let $PN:=\langle P, T, F \rangle$ be an acyclic workflow net and $\Pi$
be the set of all its paths.  Then, the set of forward reachable
states of the VAS $\langle \underline{x}, In(\underline{x}), \{
\tau_i(\underline{x}, \underline{x}') ~|~ i=1,...,n \}\rangle$
associated to $PN$ is identified by the formula $FR^{\ell}(\tau, In)$
for $\ell = max_{\pi\in\Pi} \{ len(\pi|_{T}) \}$, where $\pi|_{T}$ is
the sequence obtained from $\pi$ by forgetting each of its element in
$P$ and $len(\pi|_T)$ is the length of the sequence $\pi|_T$.

Before proving the Lemma, we instantiate the notion of symbolic
execution tree (introduced in \secref{subsec:fr-sst}) to VASs so
that we can use it in the proof of the results in this part.  The
\emph{symbolic execution tree of a VAS} is a labeled tree defined as
follows: (i) the root node is labeled by the formula $In$, (ii) a
node $n$ labeled by the formula $K$ has $d\leq n$ sons $n_1, ...,
n_d$ labeled by the formulae $Post(\tau_1, K), ..., Post(\tau_d, K)$
such that $Post(\tau_j, K)$ is satisfiable in the underlying model of
the VAS and the edge from $n$ to $n_j$ is labeled by $\tau_j$ for
$j=1, ..., d$, (iii) a node $n$ labeled by the formula $K$ has no
son, in which case $n$ is a \emph{final node}, if $Post(\tau_j, K)$ is
unsatisfiable in the underlying model of the VAS, for each $j=1, ...,
n$.  A symbolic execution tree of a VAS is \emph{$0$-complete} if it
consists of the root node labeled by the formula $In$, it is
\emph{$(d+1)$-complete} for $d\geq 0$ if its depth is $d+1$ and for
each node $n$ labeled by a formula $K_n$ at depth $d$, if
$Post(\tau_j,K_n)$ is satisfiable, then there exists a node $n'$ at
depth $d+1$ labeled by $Post(\tau_j,K_n)$.  
\begin{property}
  \label{prop:FR-vs-symbolic-execution-tree}
  Let $\langle \underline{x}, In(\underline{x}), \{
  \tau_i(\underline{x}, \underline{x}') ~|~ i=1,...,n \}\rangle$ be a
  VAS. The disjunction of all the formulae labeling a $d$-complete
  symbolic execution tree of the VAS above is logically equivalent to
  $FR^{\ell}(\tau, In)$, where $\tau:=\bigvee_{i=1}^n \tau_i$.
\end{property}
\begin{IEEEproof}
  First of all, observe that $Post$ distributes over disjunction,
  i.e.\

\vspace*{-0.4cm}
{\footnotesize
  \begin{eqnarray*}
    && Post(\tau,K)\\
    & [\mbox{by definition of } Post]~~ \Leftrightarrow & 
   \exists \underline{x}'.(K(\underline{x}') \wedge 
                           \tau_i(\underline{x}',\underline{x})) \\
    & [\mbox{by definition of } \tau]~~ \Leftrightarrow & 
   \exists \underline{x}'.(K(\underline{x}') \wedge 
                           \bigvee_{i=1}^n \tau_i(\underline{x}',\underline{x})) \\
    & [\mbox{by property of $\wedge,\vee, \exists$}]~~ \Leftrightarrow & 
   \bigvee_{i=1}^n \exists \underline{x}'.(K(\underline{x}') \wedge \tau_i(\underline{x}',\underline{x})) \\
    & [\mbox{by definition of }Post]~~ \Leftrightarrow & 
    \bigvee_{i=1}^n Post(\tau_i, K) .
  \end{eqnarray*}
}%
  Then, the property follows by a simple induction on the depth of the
  symbolic execution tree.  
\end{IEEEproof}

Interestingly, we observe that the fix-point checks are always successful
for every formula $FR^{\ell'}(\tau,In)$ with $\ell' > max_{\pi\in\Pi}
\{ len(\pi|_{T}) \}$, since $FR^{\ell'}(\tau,In)=FR^{\ell}(\tau,In)$
as no transition is enabled in $FR^{\ell}(\tau,In)$.  We can rephrase
this in terms of the symbolic execution tree as follows.
\begin{property}
  \label{prop:VAS-tree-completeness}
  Let $PN:=\langle P, T, F \rangle$ be an acyclic WF net. The
  symbolic execution tree of the VAS associated to $PN$ is
  $\ell$-complete for every $\ell\geq max_{\pi\in\Pi} \{ len(\pi|_{T})
  \}$.
\end{property}
\begin{IEEEproof}[Proof sketch]
  This is a consequence of the previous property and the observation
  that $FR^{\ell'}(\tau,In)=FR^{\ell}(\tau,In)$ for every $\ell' >
  max_{\pi\in\Pi} \{ len(\pi|_{T}) \}$.  
\end{IEEEproof}

Now, we establish a connection between (projections of) paths in a
Petri net and (projections of) paths in a symbolic execution tree.  
\begin{property}
  Let $PN:=\langle P, T, F \rangle$ be a Petri net, and $\Pi$ be the set
  of all its paths.  Then, for each $\pi\in \Pi$, there exists a path
  $\pi'$ in the symbolic execution tree of the VAS associated to $PN$
  such that $\pi|_T = \pi'|_T$.  The vice versa also holds, i.e.\ for
  each $\pi'$ in the symbolic execution tree of the VAS associated to
  $PN$, there exists a path $\pi\in \Pi$ such that $\pi|_T = \pi'|_T$.
\end{property}
\begin{IEEEproof}[Proof sketch]
  This is a consequence of the previous property and the fact that
  sets of reachable markings of Petri nets and sets of reachable
  states of associated VASs are in a one-to-one correspondence. 
\end{IEEEproof}

Lemma~\ref{lem:acyclic-workflow-net} is a consequence of the last
property above and the fact that all the paths in the net are of
bounded length so that it is possible to compute the one with maximal
length.  

\section*{Proofs of \secref{subsec:rbac4bpel-tfr}}

\paragraph*{Fact~\ref{fact:post-RBAC}}
$Post(K,\tau_i)$ is equivalent to

\vspace*{-0.4cm}
{\footnotesize
\begin{eqnarray*}
  (\exists \underline{u}.(K(\mathit{xcd}) \wedge
  \mathit{xcd}({u}_j,t) \wedge
  \xi(\underline{u},\mathit{xcd}) )) & \vee \\
  (\exists \underline{u}.(K[\lambda x,y.(\neg (x=u_j\wedge y=t) \wedge 
  \mathit{xcd}(x,y))] \wedge \\ 
  \xi[\underline{u},\lambda x,y.(\neg (x=u_j\wedge y=t) \wedge 
  \mathit{xcd}(x,y))])) & , 
\end{eqnarray*}
}%
where $K[\lambda x,y.(\neg (x=u_j\wedge y=t) \wedge
\mathit{xcd}(x,y))]$ is the formula obtained from $K$ by substituting
each occurrence of $xcd'$ with the $\lambda$-expression in the square
brackets and then performing the $\beta$-reduction and similarly for
$\xi[\underline{u},\lambda x,y.(\neg (x=u_j\wedge y=t) \wedge
\mathit{xcd}(x,y))]$.
\begin{IEEEproof}
  First of all, observe the following.  Assume $\mathit{xcd}=\lambda
  x,y.((x=u_j\wedge y=t)\vee \mathit{xcd}'(x,y))$.  We have (a)
  $\mathit{xcd}'=\mathit{xcd}$ if $\mathit{xcd}({u}_j,t)$ holds and
  (b) $\mathit{xcd}'=\lambda x,y.(\neg (x=u_j\wedge y=t)\wedge
  \mathit{xcd}(x,y)$ otherwise (i.e.\ when $\neg
  \mathit{xcd}({u}_j,t)$).  Now, consider the following simple
  transformations:

\vspace*{-0.4cm}
{\footnotesize
  \begin{eqnarray*}
  && Post(K,\tau_i) \\
  & \Leftrightarrow & \exists \mathit{xcd}'.(K(\mathit{xcd}') \wedge 
                           \tau_i(\mathit{xcd}',\mathit{xcd})) \\
  & \Leftrightarrow & \exists \mathit{xcd}'.(K(\mathit{xcd}') \wedge 
      \exists \underline{u}.(\xi(\underline{u},\mathit{xcd}') \wedge \\
  &&       \forall x,y.(\mathit{xcd}(x,y) \Leftrightarrow 
                          ((x=u_j\wedge y=t) \vee \mathit{xcd}'(x,y))))) \\
  & \Leftrightarrow & \exists \mathit{xcd}'.(K(\mathit{xcd}') \wedge \\
  && \exists \underline{u}.((\mathit{xcd}'({u}_j,t) \vee
                             \neg \mathit{xcd}'(\underline{u}_j,t) ) \wedge
      \xi(\underline{u},\mathit{xcd}') \wedge \\
  &&       \forall x,y.(\mathit{xcd}(x,y) \Leftrightarrow 
                          ((x=u_j\wedge y=t) \vee \mathit{xcd}'(x,y))))) \\
  & \Leftrightarrow & \exists \mathit{xcd}'.(K(\mathit{xcd}') \wedge \\
  && (\exists \underline{u}.((\mathit{xcd}'({u}_j,t)) \wedge
      \xi(\underline{u},\mathit{xcd}') \wedge \\
  &&       \forall x,y.(\mathit{xcd}(x,y) \Leftrightarrow 
                          ((x=u_j\wedge y=t) \vee \mathit{xcd}'(x,y))))) \vee \\
  && (\exists \underline{u}.((\neg \mathit{xcd}'({u}_j,t)) \wedge
      \xi(\underline{u},\mathit{xcd}') \wedge \\
  &&       \forall x,y.(\mathit{xcd}(x,y) \Leftrightarrow 
                          ((x=u_j\wedge y=t) \vee \mathit{xcd}'(x,y)))))) \\
  & \Leftrightarrow & \exists \mathit{xcd}'.(K(\mathit{xcd}') \wedge \\
  && (\exists \underline{u}.((\mathit{xcd}'({u}_j,t)) \wedge
      \xi(\underline{u},\mathit{xcd}') \wedge \\
  &&       \forall x,y.(\mathit{xcd}(x,y) \Leftrightarrow 
                          \mathit{xcd}'(x,y)))) \vee \\
  && (\exists \underline{u}.((\neg \mathit{xcd}'({u}_j,t)) \wedge
      \xi(\underline{u},\mathit{xcd}') \wedge \\
  &&       \forall x,y.(\mathit{xcd}(x,y) \Leftrightarrow 
                          ((x=u_j\wedge y=t) \vee \mathit{xcd}'(x,y)))))) \\
  & \Leftrightarrow & \exists \mathit{xcd}'.(K(\mathit{xcd}') \wedge \\
  && (\exists \underline{u}.((\mathit{xcd}'({u}_j,t)) \wedge
      \xi(\underline{u},\mathit{xcd}') \wedge \\
  &&       \forall x,y.(\mathit{xcd}(x,y) \Leftrightarrow 
                          \mathit{xcd}'(x,y)))) \vee \\
  && (\exists \underline{u}.((\neg \mathit{xcd}'({u}_j,t)) \wedge
      \xi(\underline{u},\mathit{xcd}') \wedge \\
  &&       \forall x,y.(\mathit{xcd}'(x,y) \Leftrightarrow 
                          (\neg (x=u_j\wedge y=t) \wedge \mathit{xcd}(x,y)))))) \\
  & \Leftrightarrow & \exists \mathit{xcd}'.( \\
  && (\exists \underline{u}.(K(\mathit{xcd}') \wedge (\mathit{xcd}'({u}_j,t)) \wedge
      \xi(\underline{u},\mathit{xcd}') \wedge \\
  &&       \forall x,y.(\mathit{xcd}(x,y) \Leftrightarrow 
                          \mathit{xcd}'(x,y)))) \vee \\
  && (\exists \underline{u}.(K(\mathit{xcd}') \wedge (\neg \mathit{xcd}'({u}_j,t) \wedge
      \xi(\underline{u},\mathit{xcd}') \wedge \\
  &&       \forall x,y.(\mathit{xcd}'(x,y) \Leftrightarrow 
                          (\neg (x=u_j\wedge y=t) \wedge \mathit{xcd}(x,y)))))) \\
  & \Leftrightarrow &  (\exists \underline{u}.(K(\mathit{xcd}) \wedge (\mathit{xcd}({u}_j,t)) \wedge
      \xi(\underline{u},\mathit{xcd}) 
)) \vee \\
  && (\exists \underline{u}.(K[\lambda x,y.(\neg (x=u_j\wedge y=t) \wedge \mathit{xcd}(x,y))] \wedge \\ 
  &&  \quad\quad\xi[\underline{u},\lambda x,y.(\neg (x=u_j\wedge y=t) \wedge \mathit{xcd}(x,y))]
                          )) .
  \end{eqnarray*}
}%
  This concludes the proof.
\end{IEEEproof}

For completeness, as done for VASs, we instantiate the notion of
symbolic execution tree (introduced in \secref{subsec:fr-sst}) to
RBAC4BPEL systems.  The \emph{symbolic execution tree of a RBAC4BPEL
  system} is a labeled tree defined as follows: (i) the root node is
labeled by the formula $In$, (ii) a node $n$ labeled by the formula
$K$ has $d\leq n$ sons $n_1, ..., n_d$ labeled by the formulae
$Post(\tau_1, K), ..., Post(\tau_d, K)$ such that $Post(\tau_j, K)$ is
satisfiable in the underlying model of the RBAC4BPEL and the edge from
$n$ to $n_j$ is labeled by $\tau_j$ for $j=1, ..., d$, (iii) a node
$n$ labeled by the formula $K$ has no son, in which case $n$ is a
\emph{final node}, if $Post(\tau_j, K)$ is unsatisfiable in the
underlying model of the RBAC4BPEL, for each $j=1, ..., n$.  A symbolic
execution tree of a RBAC4BPEL system is \emph{$0$-complete} if it
consists of the root node labeled by the formula $In$, it is
\emph{$(d+1)$-complete} for $d\geq 0$ if its depth is $d+1$ and for
each node $n$ labeled by a formula $K_n$ at depth $d$, if
$Post(\tau_j,K_n)$ is satisfiable, then there exists a node $n'$ at
depth $d+1$ labeled by $Post(\tau_j,K_n)$.
\begin{property}
  Let $\langle \underline{p}, In(\underline{p}), \{
  \tau_i(\underline{p}, \underline{p}') ~|~ i=1,...,n \}\rangle$ be a
  RBAC4BPEL system.  The disjunction of all the formulae labeling a
  $d$-complete symbolic execution tree of the RBAC4BPEL system above
  is logically equivalent to $FR^{\ell}(\tau, In)$, where
  $\tau:=\bigvee_{i=1}^n \tau_i$.
\end{property}

The proof is almost identical to that of
Property~\ref{prop:FR-vs-symbolic-execution-tree} and it is thus
omitted.

\section*{Proofs of \secref{subsec:combining-vas-RBAC4BEPL}}

Preliminarily, we modularly define the sequence of formulae
characterizing sets of forward reachable states and the symbolic
execution trees of VAS+RBAC4BPEL systems by re-using the associated
VAS and RBAC4BPEL system.

For the formulae describing the set of forward reachable states,
define the following sequence, by recursion: $FR^0(K,\tau)
:= K$ and $FR^{i+1}(K,\tau):=(Post_V^i(FR^i,\tau)\wedge
Post_R^i(FR^i,\tau))\vee FR^i(K,\tau)$ for $i\geq 0$, $K:=K_V\wedge
K_R$, $K_V$ is a formula of LA, $K_R$ is a BSR formula, and
$\tau:=\bigvee_{k=1}^n \tau_i$.  

For the \emph{symbolic execution system of a VAS+RBAC4BPEL system},
preliminarily introduce the following two labeling functions.  Given a
node of the symbolic execution tree for a VAS+RBAC4BPEL system, the
\emph{VAS-labeling} function returns $Post_V^i(FR^i,\tau)$ while the
\emph{RBAC4BPEL-labeling} function returns $Post_R^i(FR^i,\tau)$.
Then, the \emph{symbolic execution tree of a VAS+RBAC4BPEL system} is
a (multi-)labeled tree defined as follows: (i) the root node is
VAS-labeled by the formula $In_V$ and RBAC4BPEL-labeled by the
formula $In_R$, (ii) a node $n$ VAS-labeled by the formula $K_V$ and
RBAC4BPEL-labeled by the formula $K_R$ has $d\leq n$ sons $n_1, ...,
n_d$, each $n_j$ is VAS-labeled by the formula $Post_V(\tau_j, K)$
such that $Post_V(\tau_j, K)$ is satisfiable in the structure
underlying the associated VAS and it is RBAC4BPEL-labeled by the
formula $Post_R(\tau_j, K)$ such that $Post_R(\tau_j, K)$ is
satisfiable in the structure underlying the associated RBAC4BPEL
system, and the edge from $n$ to $n_j$ is labeled by $\tau_j$ for
$j=1, ..., d$, (iii) a node $n$ labeled by the formula $K$ has no
son, in which case $n$ is a \emph{final node}, if both $Post_V(\tau_j,
K)$ is unsatisfiable modulo Linear Arithmetic and $Post_R(\tau_j, K)$
is unsatisfiable modulo the BSR theory, for each $j=1, ..., n$.  A
symbolic execution tree of a VAS+RBAC4BPEL system is
\emph{$0$-complete} if it consists of the root node VAS-labeled by
the formula $In_V$ and RBAC-labeled by the formula $In_R$, it is
\emph{$(d+1)$-complete} for $d\geq 0$ if its depth is $d+1$ and for
each node $n$ labeled by a formula $K_n$ at depth $d$, if both
$Post_V(\tau_j,K_n)$ is satisfiable modulo LA and $Post_R(\tau_j,K_n)$
is satisfiable modulo the BSR theory, then there exists a node $n'$ at
depth $d+1$ VAS-labeled by $Post_V(\tau_j,K_n)$ and
RBAC4BPEL-labeled by $Post_R(\tau_j,K_n)$.


\begin{property}
  \label{prop:FR-vs-symbolic-execution-tree-VAS+RBAC}
  Let $\langle \underline{x},\underline{p}, In_V(\underline{x})\wedge
  In_R(\underline{p}), \{ \tau_i^V(\underline{x},\underline{x}')\wedge
  \tau_i^R(\underline{p},\underline{p}')~|~ i=1, ..., n\} \rangle$ be
  a VAS+RBAC system.  The disjunction of the conjunction between the
  VAS-labeling and RBAC-labeling formulae of all the nodes in a
  $d$-complete symbolic execution tree is logically equivalent to
  $FR^{\ell}(\tau,In)$, where $\tau:=\bigvee_{i=1}^n \tau_i$.
\end{property}

The proof is along the lines of that of
Property~\ref{prop:FR-vs-symbolic-execution-tree} and uses the
modularity of the post-image computation (see
Fact~\ref{fact:modular-post-image}).

We are now ready to prove the main result of the paper.
\paragraph*{Theorem~\ref{th:main}}
Let $PN:=\langle P,T,F \rangle$ be a an acyclic WF net and
$\langle \underline{x}, In_V(\underline{x}), \{
\tau_i^V(\underline{x},\underline{x}') ~|~ i=1, ..., n\} \rangle$ be
its associated VAS.  Furthermore, let $\langle \underline{p},
In_R(\underline{p}), \{ \tau_i^R(\underline{p},\underline{p}') ~|~
i=1, ..., n\} \rangle$ be the RBAC system with a finite and known set
of users.  Then, the symbolic reachability problem of the
VAS+RBAC4BPEL system whose associated VAS and RBAC4BPEL system are
those specified above is decidable.
\begin{IEEEproof}
  Let $G_V(\underline{x})\wedge G_R(\underline{p})$ be a goal formula
  such that $G_V$ is a LA formula and $G_R$ be a BSR formula.  By
  Property~\ref{prop:finiteness-symb-sim-tree-VAS+RBAC}, we know that
  there exists a bound $\overline{\ell}\geq 0$ such that for every
  $\ell\geq \overline{\ell}$, the symbolic execution tree of the
  VAS+RBAC4BPEL system is $\overline{\ell}$-complete.  Furthermore, by
  Property~\ref{prop:FR-vs-symbolic-execution-tree-VAS+RBAC}, we know
  that the disjunction of all VAS-labeling and RBAC4BPEL-labeling
  formulae is equivalent to $FR^{\overline{\ell}}(\tau,In)$.  Because
  of the $\overline{\ell}$-completeness of the symbolic execution
  tree, we know that $FR^{\overline{\ell}}(\tau,In)$ is a fix-point;
  hence, $FR^{\overline{\ell}}(\tau,In)$ describes the set of all
  forward reachable states of the VAS+RBAC4BPEL system.  By induction
  on the length of the sequence $FR^0,FR^1, ...$ of formulae, it is
  easy to show that each $FR^i$ is equivalent to the conjunction
  between a formula of Linear Arithmetic, say $FR^i_V$, and a BSR
  formula, say $FR^i_R$.  Hence, we conclude that
  $FR^{\overline{\ell}}(\tau,In)$ is equivalent to
  $FR^{\overline{\ell}}_V\wedge FR^{\overline{\ell}}_R$.  Thus, in
  order to solve the goal reachability problem, it is sufficient to
  check the satisfiability of the following formula:

\vspace*{-0.4cm}
{\footnotesize
  \begin{eqnarray*}
    (FR^{\overline{\ell}}_V\wedge FR^{\overline{\ell}}_R) & \wedge &
    (G_V(\underline{x})\wedge G_R(\underline{p}))
  \end{eqnarray*}
}%
  modulo the union of LA and the BSR theory.  This problem can be
  reduced to two separate satisfiability problems modulo a single
  theory, namely: (i) checking the satisfiability of

\vspace*{-0.4cm}
{\footnotesize
  \begin{eqnarray*}
    FR^{\overline{\ell}}_V &\wedge& G_V(\underline{x})
  \end{eqnarray*}
}%
  modulo Linear Arithmetic and (ii) checking the satisfiability of 

\vspace*{-0.4cm}
{\footnotesize
  \begin{eqnarray*}
    FR^{\overline{\ell}}_R & \wedge & G_R(\underline{p})
  \end{eqnarray*}
}%
  modulo the BSR theory.  Both of these problems are well-known to be
  decidable and hence the overall problem is decidable.  This
  concludes the proof.
\end{IEEEproof}
\end{LONG}

\end{document}